\documentclass[prl,aps,showpacs,preprint]{revtex4}
\usepackage{epsfig}
\begin{document}

\title{Anticipating the response of excitable systems driven by random forcing}

\author{M. Ciszak$^1$, O. Calvo$^1$, C. Masoller$^{1,3}$, Claudio R. Mirasso$^1$ and Ra\'{u}l Toral$^{1,2}$}

\affiliation{$^1$ Departament de F\'{\i}sica, Universitat de les Illes
Balears, E-07071 Palma de Mallorca, Spain\\
$^2$ Instituto Mediterraneo de Estudios Avanzados,
CSIC-UIB, E-07071 Palma de Mallorca, Spain\\
$^3$Instituto de F\'\i sica, Facultad de Ciencias,
Universidad de la Rep\'ublica, Igua 4225, Montevideo 11400,
Uruguay}

\date{\today}

\begin{abstract}
We study the regime of anticipated synchronization in unidirectionally coupled model neurons subject to a common external aperiodic forcing that makes their behavior unpredictable. We show numerically and by implementation in analog hardware electronic circuits that, under appropriate coupling conditions, the pulses fired by the slave neuron anticipate (i.e. predict) the pulses fired by the master neuron. This anticipated synchronization occurs even when the common external forcing is white noise.
\end{abstract}

\maketitle


Synchronization of nonlinear systems is a fascinating subject that has been extensively studied on a large variety of physical and biological systems\cite{PRK01}. While synchronization of oscillators goes back to the work by Huygens, the last decade has witnessed an increased interest in the topic of synchronization of chaotic systems \cite{B02}. 

Recently, Voss \cite{VOSS} has discovered a new scheme of synchronization, called ``anticipated synchronization''. Voss has shown that by using appropriate delay lines it is possible to synchronize two unidirectionally coupled systems in such a way that the {\sl slave} system, ${\bf y}(t)$, predicts the behavior of the {\sl master} system, ${\bf x}(t)$. Two coupling schemes were considered: {\sl complete replacement},
\begin{equation}
\label{scheme1}
\begin{array}{rcl}
\dot {\bf x}(t)& =& -\alpha {\bf x}(t)+{\bf f}({\bf x}(t-\tau))\\
\dot {\bf y}(t) & = & -\alpha {\bf y}(t)+{\bf f}({\bf x}(t))
\end{array}
\end{equation}
and {\sl delay coupling}, 
\begin{equation}
\label{scheme2}
\begin{array}{rcl}
\dot {\bf x}(t)& =& {\bf f}({\bf x}(t))\\
\dot {\bf y}(t) & = & {\bf f}({\bf y}(t))+{\bf K}[{\bf x}(t)-{\bf y}(t-\tau)].
\end{array}
\end{equation}
${\bf f}({\bf x})$ is a function which defines the
{\it autonomous} dynamical system under consideration, ${\bf K}$ is the coupling strength and $\tau$ is a delay time. It is easy to see that in both schemes the manifold \mbox{${\bf y}(t)={\bf x}(t+\tau)$} is a solution of the equations, and Voss has shown that in both schemes this solution can be structurally stable. This is more remarkable when the dynamics of the master system $\bf x$ is ``intrinsically unpredictable", as it is the case of a chaotic system. While in the scheme of complete replacement the anticipation time $\tau$ can be arbitrarily large, the delay coupling scheme requires some constrains on $\tau$ and $\bf K$ for the synchronization solution to be stable \cite{VOSS}. Despite this fact, the delay coupling scheme is more interesting since the anticipation time $\tau$ can be varied without altering the dynamics of the master system $\bf x$. 

In this Letter we study numerically and experimentally the regime of anticipated synchronization in excitable {\it non-autonomous} systems. In our case the intrinsic unpredictability of the behavior of the dynamical system $\bf x$ does not arise from a chaotic dynamics, but rather from the existence of an external forcing with some element of randomness. We consider the coupled systems
\begin{equation}
\begin{array}{rcl}
\dot {\bf x}(t)& =& {\bf f}({\bf x}(t))+{\bf I}(t)\\
\dot {\bf y}(t) & = & {\bf f}({\bf y}(t))+{\bf I}(t)+{\bf K}[{\bf x}(t)-{\bf y}(t-\tau)] \label{eq3},
\end{array}
\end{equation}
where ${\bf I}(t)$ represents a common external forcing. Notice that  ${\bf y}(t)={\bf x}(t+\tau)$ is no longer an exact solution of the equations [except in the particular case of a periodic forcing ${\bf I}(t+\tau)={\bf I}(t)$]. We show that under appropriate coupling conditions there can be a very good correlation between ${\bf y}(t)$ and ${\bf x}(t+\tau)$ which, in practice, allows the prediction of the future behavior of ${\bf x}(t)$ with a high degree of accuracy. Of course, this result is more remarkable when the external forcing is a random signal.

Specifically, we have considered models of sensory neurons.
Sensory neurons transform external stimuli signals as pressure, temperature,
electric pulses, etc., into trains of action potentials, usually referred to as
'spikes' or 'firings'. Their behavior is typical of excitable systems: if the
forcing is above a certain threshold, the neuron fires a pulse, and
after the firing, the recovery process produces an absolute refractory time
during which a second firing cannot occur. 

In general, sensory neurons work in a noisy environment. As a consequence, the time intervals between spikes contain a significant random component, and random spikes often occur even in the absence of stimuli. The topics of synchronous oscillations and noise have received much attention (see,
e.g., \cite{LBM91}), since it has been suggested that synchronous firing activity of sensory neurons might be a part of higher brain functions and a method for integrating distributed information into a global picture \cite{GM98}.

Here we find that the interplay of coupling, delayed feedback, and common noise can lead to anticipated synchronization. We illustrate this effect in the well known FitzHugh-Nagumo and Hodkey-Huxley neuron models. By coupling two of such systems in an unidirectional configuration as in the scheme (\ref{eq3}), we find that when both systems are subjected to the same external random forcing, the slave system fires the same train of spikes as the master system, but at a certain amount of time earlier, i.e., the slave predicts the response of the master. 

\begin{figure}
\centerline{\epsfig{file=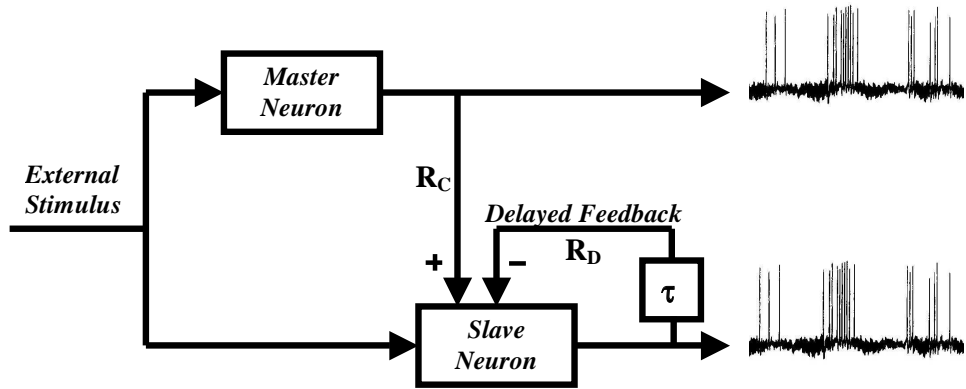,width=.9\textwidth,angle=-0}}
\caption{\label{fig1} Schematic diagram of two model neurons coupled in a unidirectional configuration, subjected to the same external forcing and with a feedback loop (with a delay time $\tau$) in the slave neuron.
}
\end{figure}

First we show results based on the FitzHugh-Nagumo model. It consists of two variables ${\bf x}=(x_1,x_2)$. The fast variable, $x_1$, is associated with the activator, and the slow recovery variable, $x_2$, is associated with the inhibitor. The equations for the master $(x_1, x_2)$ and the slave ${\bf y}=(y_1, y_2)$ systems, under unidirectional coupling are, respectively (see the schematic diagram shown in Fig. \ref{fig1}):
\begin{equation}
\label{eq1}
\begin{array}{rcl}
\dot{x_1} & = & -x_1(x_1-a)(x_1-1)-x_2 + I(t)\\
\dot{x_2}& = & \epsilon (x_1-b x_2)
\end{array}
\end{equation}
and
\begin{equation}
\label{eq2}
\begin{array}{rcl}
\dot{y_1} & = & -y_1(y_1-a)(y_1-1)-y_2 + I(t)+ \\ & & + K[x_1(t)-y_1(t-\tau)]\\
\dot{y_2}& = & \epsilon (y_1-b y_2)
\end{array}
\end{equation}
where $a$, $b$, and $\epsilon$ are constants, $K$ is the coupling
strength and $\tau$ is a delay time (associated to an inhibitory feedback loop in the slave neuron). Note that only the fast variables of the two systems are coupled. When the common external forcing, $I(t)$, is constant in time, the anticipated synchronization manifold: $x_1(t+\tau)=y_1(t)$, $x_2(t+\tau)=y_2(t)$ is an exact solution of Eqs. (\ref{eq1}) and (\ref{eq2}). If the external forcing is above threshold and for appropriate values of $K$ and $\tau$, the master system fires pulses periodically and the coupling induces a constant time shift $\tau$ between master and slave spikes.

\begin{figure}
\centerline{\epsfig{file=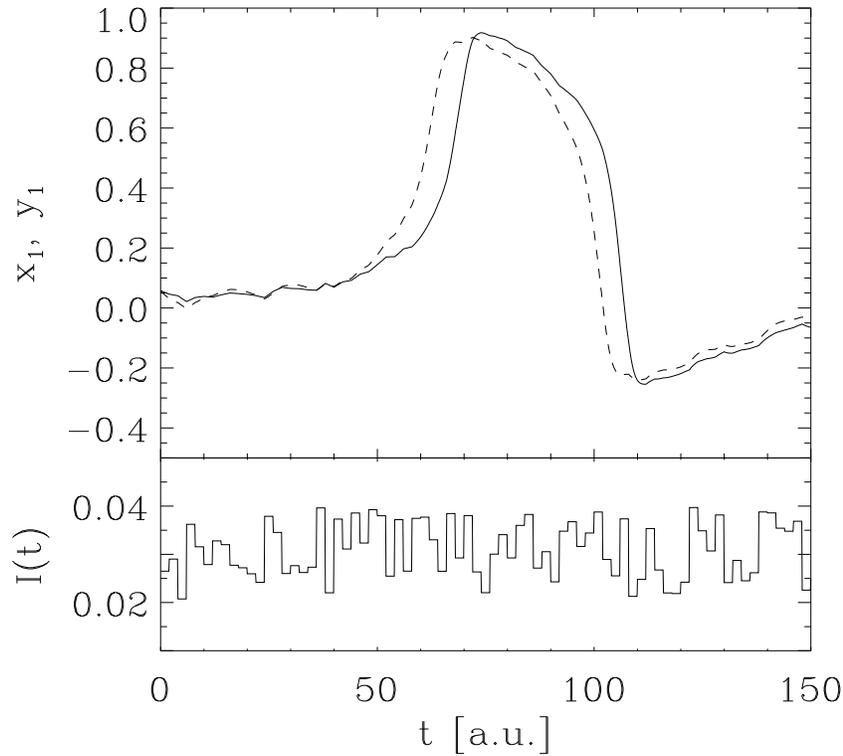,width=.75\textwidth,angle=-0}}
\caption{\label{fig2}Anticipated synchronization from a numerical integration of the FitzHugh-Nagumo set of Eqs.(\ref{eq1}-\ref{eq2}). The parameters are: $a=0.139$, $b=2.54$, $\epsilon=0.008$, $K=0.15$. The external forcing $I(t)$ (displayed in the lower panel) is a random amplitude noise of period $T=2$, mean value $I_0=0.03$ and amplitude $D=0.01$. Notice (upper panel) that the pulse of the slave system $y_1(t)$ (dashed line) anticipates the pulse of the master system $x_1(t)$ (solid line) by a time approximately equal to the time delay $\tau=4$.}
\end{figure}

We have considered different types of random external forcing $I(t)$. The first
one corresponds to a random process whose amplitude remains constant for a time
$T$ and then it switches to a new random value chosen uniformly in
[$I_{0}-D$,$I_{0}+D$], where $D$ is the noise intensity. We chose $I_{0}$ very close to (but below) the firing threshold of the excitable system. It would appear at first thought that with this type of external forcing the behavior of the master system can be easily predictable. However, there are two main factors that make the system response unpredictable: if the effect of the perturbation is not strong enough the system does not fire a pulse; besides, the system has a refractory time (after firing a pulse) during which, another firing is not possible. Figure \ref{fig2} shows that anticipation occurs with this type of random external forcing for an appropriate value of the coupling strength $K$: after an initial transient time the two systems synchronize such that the slave system anticipates the fires of the master system by a time interval $\tau$. The firings in the master and the slave systems start at about the same time, and the anticipation phenomenon grows during the rising of the pulse. When the master system noisily evolves near the stable point, the anticipation vanishes. In other words, anticipation is a local process, during firings.

\begin{figure}
\centerline{\epsfig{file=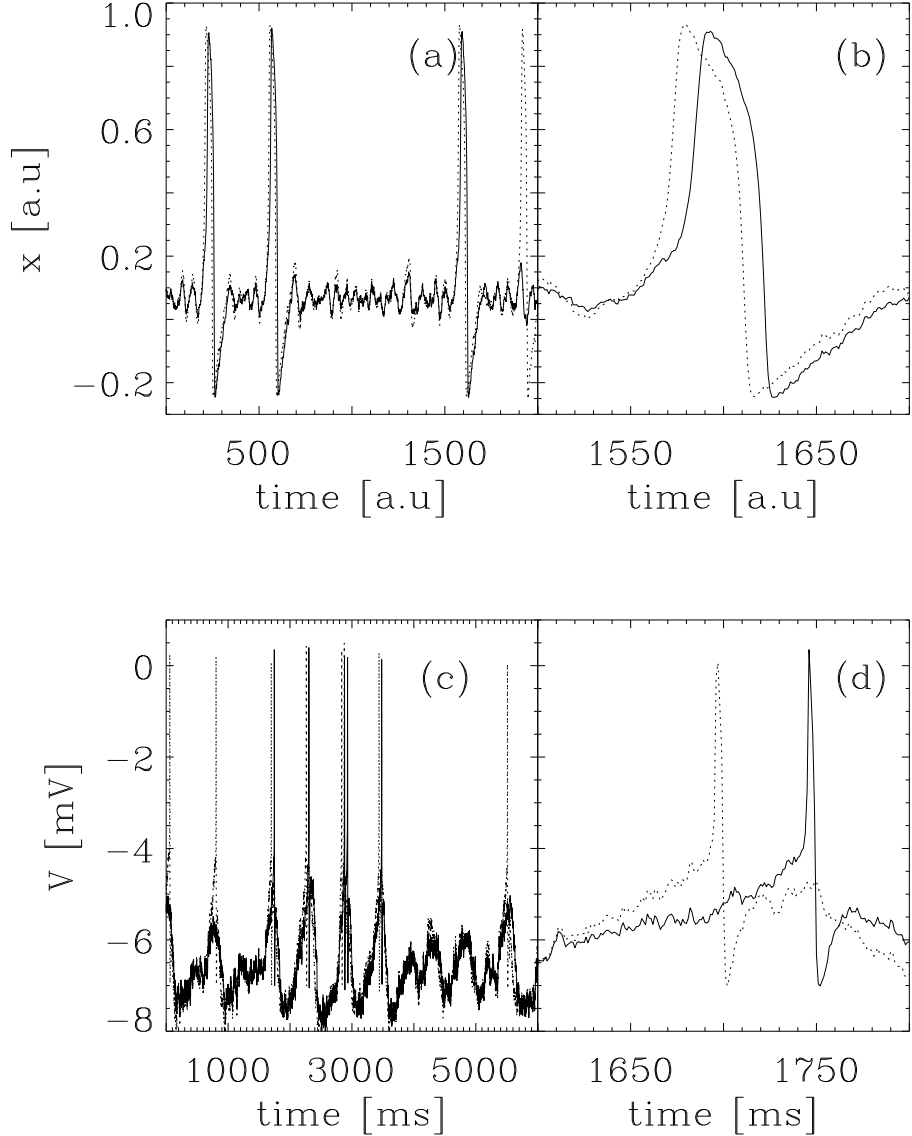,width=.75\textwidth,angle=-0}}
\caption{\label{fig3} Trains of spikes obtained from numerical simulations of models of unidirectionally coupled neurons subjected to the same external forcing, which is a Gaussian white noise with mean $I_0$ and correlations $\langle [I(t)-I_0][I(t')-I_0]\rangle =2D\delta(t-t')$. (a) Simulation of two FitzHugh-Nagumo neurons, Eqs. (\ref{eq1}-\ref{eq2}). The parameters are $a=0.139$, $b=2.54$, $\epsilon=0.008$, $I_0=0.03$, $K=0.03$, $\tau=10$, $D=2.45\times10^{-5}$. (b) Simulation of two Hodgkin-Huxley neurons, Eqs. (\ref{hh}) with $K=0.03$ ms$^{-1}$, $\tau$=50 ms, and $D$=0.5 mV$^2$/ms; all other parameters as in \cite{B98} ($T=6$ C, $V_l=-75$ mA in the notation of that paper). Left panels show typical spike trains; right panels show with detail a single spike. The solid (dashed) line represents the output of the master (slave) system.
}
\end{figure}

The same qualitative results are found with other types of external forcing such as colored or even white noise. Figures \ref{fig3}(a-b) display the spikes of the master and slave systems when $I(t)$ is Gaussian white noise.
  
Sometimes the slave system makes an error in anticipating the master firings.
While the slave system always fires a pulse when the master system fires a
pulse, it also might fire a ``extra'' pulse, which has no corresponding pulse
in the train of pulses fired by the master.  Notice that in Fig. \ref{fig3}(a) 
an error at about $t=1900$ occurs. Not surprisingly, we find that
the longer the anticipation time $\tau$, the larger the number of errors.
However, for a given anticipation time, the number of errors can be reduced
considerably if a ``cascade'' of an adequate number of slave neurons is
considered. A detailed study of the number of errors and its dependency with
the type of external forcing will be reported elsewhere.

Next we show simulations based on a more realistic model, namely the model of
electro-receptors proposed by Braun et. al \cite{B98}. This model is a
modification of the Hodgkin-Huxley  neuron model: \mbox{$C_M \dot{x} = 
-i_{Na}-i_K-i_{sd}-i_{sr}-i_l$}, where $x$ is the potential voltage across
the membrane and $C_M$ is the capacitance; $i_{Na}$ and $i_K$ are the fast
sodium and potassium currents, $i_{sd}$ and $i_{sr}$ are additional slow
currents, $i_l$ is a passive leak current.
For details and functional dependence of the currents on the voltage $x$ and other factors (as temperature) see \cite{B98}.

We extend the model to account for two unidirectionally coupled neurons,
with a delayed feedback loop in the slave neuron, and subject to a common external forcing $I(t)$, in the same way as in the FitzHugh-Nagumo
model, e.g., the equations for the master, $x$, and for the slave, $y$, neurons are:
\begin{equation}
\label{hh}
\begin{array}{rcl}
C_M \dot x & = & -i_{Na}^x-i_{K}^x -i_{sd}^x-i_{sr}^x-i_l^x+I(t) \\
C_M \dot y & = & -i_{Na}^y-i_{K}^y -i_{sd}^y-i_{sr}^y-i_l^y+I(t) \nonumber \\
& + & K[x(t)-y(t-\tau)]~~.
\end{array}
\end{equation}
Figures \ref{fig3}(c-d) display the results when the common external forcing $I(t)$ is a Gaussian white noise. We chose parameters such that in the absence of forcing there are no spikes (subthreshold, noise-activated firing regime). The behavior observed is qualitatively the same as in the FitzHugh-Nagumo model (the slave neuron anticipates the fires of the master neuron), which indicates that the anticipation phenomenon is general and model independent. Remarkably, in this model the anticipation time can be larger than the pulse duration. It is worth mentioning that anticipated synchronization is also observed in this model for parameters such that there is spontaneous (regular or irregular) spike activity (suprathreshold firing regime).

\begin{figure}
\centerline{\epsfig{file=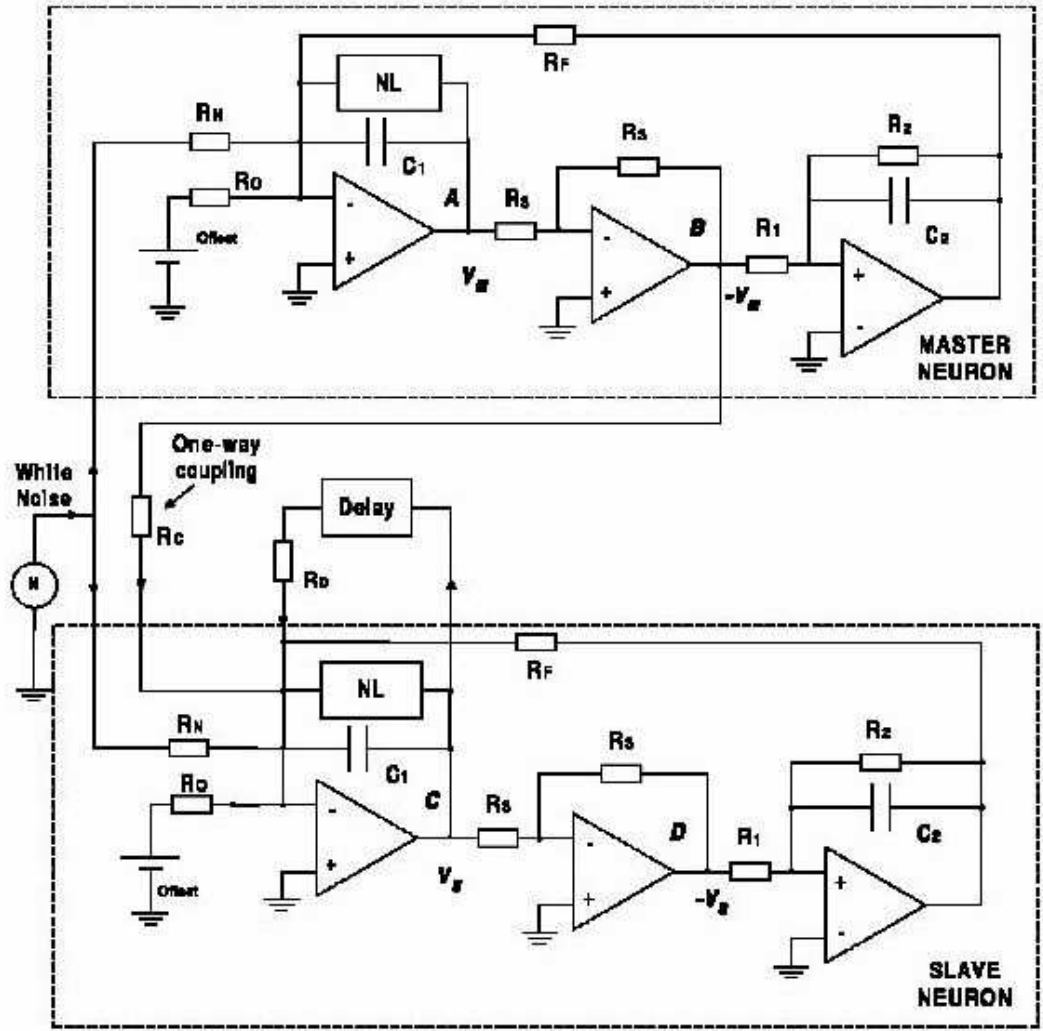,width=.9\textwidth,angle=-0}}
\caption{\label{fig4}
Circuit implementation of two coupled neurons. $R_1$ = 125 k$\Omega$, $R_2$=50 k$\Omega$, $R_3$=10 k$\Omega$, $R_C=R_D$=100 k$\Omega$, $R_F$=10 k$\Omega$, $R_N$=10 k$\Omega$, $R_O$= 10 k$\Omega$, $C_1$=100 nF, $C_2$=1 $\mu$F.
}
\end{figure}

To assess the robustness of the anticipated synchronization observed in the numerical simulations, we have implemented the FitzHugh-Nagumo model in analog hardware and constructed two coupled electronic neurons (a simplified version of the circuit is shown in Fig. \ref{fig4}).
The electronic neurons were built using operational amplifiers and the cubic non-linearity described by $x (x-a) (x-1)$ was implemented using analog multipliers (AD633) in a circuit not shown for simplicity. The resistor $R_C$ controls the strength of the unidirectional coupling between the master and the slave neurons. The resistor $R_D$ ($R_D=R_C$ in our case) controls the strength of the delayed feedback into the slave neuron. The coupling and the delayed feedback have opposite signs: while the master signal was obtained at point B of Fig. \ref{fig4}, where the voltage is $-V_m$, the slave signal that goes into the delay line was obtained at point C, where the voltage is $+V_s$. The different signs are due to the inverters that are located in between points A and B and C and D. The threshold on both neurons was controlled by a potentiometer represented by its equivalent circuit: offset and $R_0$. The analog delay line for the delayed feedback in the slave neuron was built using bucket brigade circuits (MN3004). A function generator with white noise output capabilities (HP33120A) was used to excite both electronic neurons. The signals were acquired using LabView and National Instruments DAQ 6025E data acquisition board. 

Similar electronic neurons have been implemented in \cite{abarbanel}, where it was shown that their behavior is very similar to that of biological neurons: when interfaced to biological neurons, hybrid circuits, with the electronic neurons taking the place of missing or damaged biological neurons, could function normally. 

\begin{figure}
\centerline{\epsfig{file=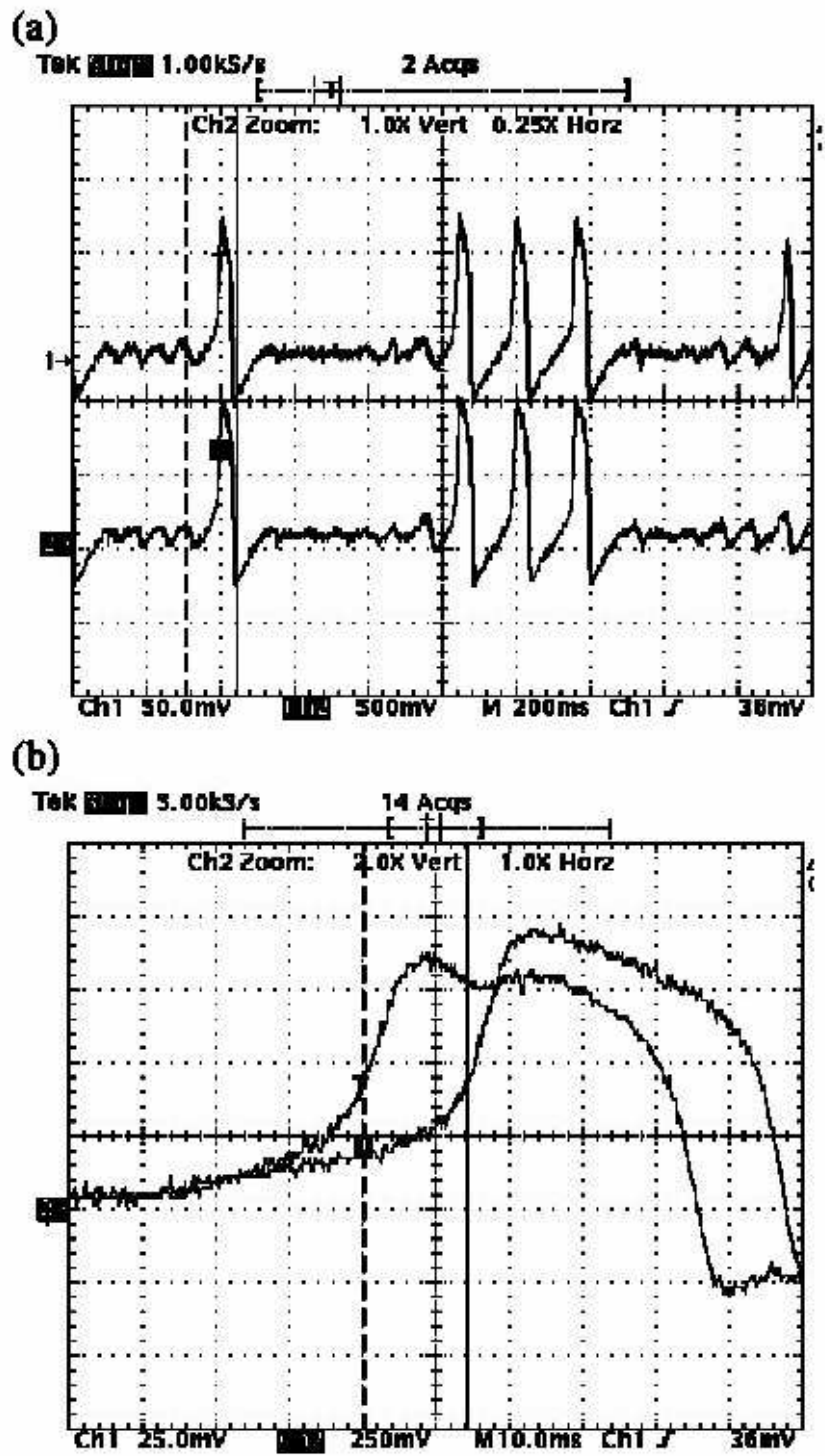,width=.9\textwidth,angle=-0}}
\caption{\label{fig5}(a) Experimental train of spikes that shows anticipation in the spikes fired by the slave neuron (upper trace) with respect to the spikes fired by the master neuron (lower trace). (b) Detail of a spike fired by the master neuron and anticipated spike fired by the slave neuron. The anticipation time is $14$ ms approximately.}
\end{figure}

Our electronic coupled neurons behave very similar as in the numerical simulations. For an appropriate value of the coupling resistance $R_C$, we observe that, after a transient, the master and slave electronic neurons synchronize in such a way that the slave neuron anticipates the fires of the master neuron by a time interval approximately equal to the delay time $\tau$ of the feedback mechanism. Figure \ref{fig5} (a) shows a typical spike train, and Fig. \ref{fig5}(b) displays in detail a single spike \cite{movies}. We observe that, as in the numerical simulations, the firings of the master and the slave neurons start at about the same time: anticipation begins during the rising of the peak and it vanishes when the neurons are in the unexcited state. Without coupling and feedback ($R_C=R_D=0$) the neurons fire pulses which are, in general, desynchronized (due to the small mismatch between the circuits).

To summarize, we have studied the regime of anticipated synchronization in coupled systems exhibiting neuronal-type excitable behavior, when they are driven by common external aperiodic forcing. We have shown that under
appropriate conditions, the slave system can anticipate the random
spikes of the master system. This is despite of the fact that the anticipated synchronization manifold is not a solution of the equations. We have simulated numerically the coupled neurons with the FitzHugh-Nagumo and a modified Hodgkin-Huxley models and we have considered different types of random forcing. The FitzHugh-Nagumo model was also implemented in analog hardware, showing that the anticipation phenomenon is very general and robust. 

Our results show that non-linearity, noise and delayed feedback might conspire to produce new interesting and unexpected phenomena, and we hope that our findings will stimulate the search for anticipated synchronization in biological systems. 

The work is supported by MCyT (Spain) and FEDER, projects BFM2001-0341-C02-01 and BMF2000-1108. C. Masoller acknowledges partial support from the UIB.

\pagebreak

\pagebreak

\pagebreak

\pagebreak

\pagebreak

\end{document}